\documentclass[superscriptaddress,twocolumn,preprintnumbers,amsmath,amssymb,aps,prl]{revtex4-2}

\usepackage{graphicx}
\usepackage{dcolumn}
\usepackage{bm}
\usepackage{hyperref}
\usepackage[mathlines]{lineno}
\usepackage{color}
\usepackage[normalem]{ulem}
\usepackage{rotating}

\begin{document}

\title{Liquid-liquid transition in a Bose fluid near collapse}

\author{Saverio Moroni}
\email{moroni@democritos.it}
\affiliation{CNR-IOM DEMOCRITOS, Istituto Officina dei Materiali and SISSA Scuola Internazionale Superiore di Studi Avanzati, Via Bonomea 265, I-34136 Trieste, Italy}

\author{Fabio Cinti}
\email{fabio.cinti@unifi.it}
\affiliation{Dipartimento di Fisica e Astronomia, Universit\`a di Firenze, I-50019, Sesto Fiorentino (FI), Italy}
\affiliation{INFN, Sezione di Firenze, I-50019, Sesto Fiorentino (FI), Italy}
\affiliation{Department of Physics, University of Johannesburg, P.O. Box 524, Auckland Park 2006, South Africa}

\author{Massimo Boninsegni}
\email{m.boninsegni@ualberta.ca}
\affiliation{Department of Physics, University of Alberta, Edmonton, Alberta, Canada T6G 2H5}

\author{Giuseppe Pellicane}
\email{giuseppe.pellicane@unime.it}
\affiliation{Dipartimento di Scienze Biomediche, Odontoiatriche e delle Immagini Morfologiche e Funzionali, Universit\`a degli Studi di Messina, I-98125, Messina, Italy}
\affiliation{CNR-IPCF, Viale F. Stagno d'Alcontres, 37-98158, Messina, Italy}
\affiliation{School of Chemistry and Physics, University of Kwazulu-Natal, 3209 Pietermaritzburg, South Africa\\$^7$National Institute of Theoretical and Computational Sciences (NIThECS), 3209 Pietermaritzburg, South Africa}

\author{Santi Prestipino}
\email{sprestipino@unime.it}
\affiliation{Dipartimento di Scienze Matematiche e Informatiche, Scienze Fisiche e Scienze della Terra, Universit\`a degli Studi di Messina, viale F. Stagno d'Alcontres 31, I-98166, Messina, Italy}

\date{\today}

\begin{abstract}
Discovering novel emergent behavior in quantum many-body systems is a main objective of contemporary research.
In this paper, we explore the effects on phases and phase transitions of the proximity to a Ruelle-Fisher instability, marking the transition to a collapsed state.
To accomplish this, we study by quantum Monte Carlo simulations a two-dimensional system of soft-core bosons interacting through an isotropic finite-ranged attraction, with a parameter $\eta$ describing its strength.
If $\eta$ exceeds a characteristic value $\eta_c$, the thermodynamic limit is lost, as the system becomes unstable against collapse.
We investigate the phase diagram of the model for $\eta\lesssim\eta_c$,  finding --- in addition to a liquid-vapor transition --- a first-order transition between two liquid phases.
Upon cooling, the high-density liquid turns superfluid, possibly above the vapor-liquid-liquid triple temperature.
As $\eta$ approaches $\eta_c$, the stability region of the high-density liquid is shifted to increasingly higher densities, a behavior at variance with  distinguishable quantum or classical particles.
Finally, for $\eta$ larger than $\eta_c$ our simulations yield evidence of collapse of the low-temperature fluid for any density;
the collapsed system forms a circular cluster whose radius is insensitive to the number of particles.

\end{abstract}
\maketitle

A first-order transition between two liquid phases is an uncommon, still elusive phenomenon that challenges our understanding of the fluid state of matter~\cite{Tanaka2020}.
In pure systems, a liquid-liquid phase transition (LLPT) is found in tetrahedral liquids~\cite{Poole1992,Vasisht2011}, where, however, it falls in the supercooled region.
A LLPT may occur in parallel with a change in the chemical nature of the constituent particles, like in hydrogen~\cite{Pierleoni2016}, where a molecular liquid is transformed under pressure into an atomic liquid.
In equilibrium, a LLPT has been observed in phosphorus~\cite{Katayama2000,Monaco2003} and in sulfur~\cite{Henry2020}, between liquids characterized by a different degree of polymerization.
There also exist (controversial) examples of LLPT in complex molecular fluids (e.g., in triphenyl phosphite~\cite{Tanaka2004}).
The situation is clearer in models, where a genuine LLPT occurs in classical particles interacting through isotropic core-softened potentials~\cite{Hemmer1970,Jagla2001,Franzese2001} or anisotropic potentials~\cite{Molinero2006,Smallenburg2014,
Debenedetti2020}.
The mechanism commonly invoked for the onset of a LLPT is the existence of two distinct repulsive length scales in the effective interparticle potential.

We here introduce a new paradigm of LLPT with no classical counterpart --- that is, a structural transition not involving a change in the elementary constituents and/or interactions.
To this aim, we push our system (a bosonic fluid) close to its stability threshold, such as existing for particles that interact via a finite repulsion augmented with a strong enough attraction.
We will highlight the non-trivial role of quantum indistinguishability, without which the LLPT simply vanishes.
In the same system, we also document a first-order transition from liquid to superfluid, a possibility which has remained unexplored so far.

While bare interatomic forces are strongly repulsive at short distances, an effective step-like repulsion can be induced at larger separations, in the nanometer to micrometer range.
This is achieved, for instance, in ultracold bosonic gases, by means of a weak laser coupling of the atomic ground state to a highly excited Rydberg state~\cite{Henkel2010,Zeiher2016}.
Such ``soft-core'' bosons are an ideal playground for the study of supersolidity~\cite{Boninsegni2012b,Cinti2014a,PhysRevLett.132.026001}, which in this kind of systems is promoted by cluster-crystal ordering of particles at high density~\cite{Pomeau1994,Cinti2010,Saccani2011,Prestipino2019}.

We consider the scenario (which could become feasible in future experimental protocols) in which a finite-range attraction of tunable strength is added to a soft-core repulsion.
As the strength of attraction grows a Bose fluid eventually undergoes the collapsing transition predicted a long time ago by Ruelle and Fisher~\cite{Fisher1966,Ruelle1969} and observed in classical fluids~\cite{Heyes2007,Fantoni2011,Malescio2015,Prestipino2016}.

For a stable interaction (see p.~33 of Ref.~\onlinecite{Ruelle1969}), the grand-canonical partition function cannot grow faster than $\textrm{exp}(cV)$ as a function of the system volume $V$, where $c$ depends on the temperature and the chemical potential.
When stability is violated, the grand partition function is instead divergent, even for finite $V$, while a system with a fixed number $N$ of particles collapses to a compact cluster or blob with a potential energy proportional to $N^2$.
Ruelle and Fisher derived analytic criteria~\cite{Fisher1966,Ruelle1969} to ascertain whether a bounded potential with an attractive component leads to collapse in a classical system.
For a large class of regular potentials, they also proved~\cite{Fisher1966} that classical instability and quantum instability are reciprocally implied in the case of bosons.
However, the 
behavior of a quantum fluid near and beyond the stability threshold is largely unexplored, and 
can certainly be elucidated by numerical simulation.
From the experimental standpoint, the 
physics of a quantum system near collapse is relevant, for example, to cold dipolar assemblies~\cite{Koch2008,Lahaye2008}.

In this letter, we present the results of a numerical investigation of a two-dimensional (2D) system of identical particles of spin zero, hence obeying Bose statistics, interacting via a double-Gaussian (DG) potential~\cite{Malescio2015}:
\begin{equation}
u(r)=\epsilon\left[e^{-(r/\sigma)^2}-\eta e^{-(r/\sigma-3)^2}\right]
\label{eq1}
\end{equation}
with $\eta>0$ (we also set a cut-off distance $r_c=6\sigma$, beyond which $u=0$ in Eq.~\ref{eq1}).
This potential is known to characterize a fluid near collapse; it has the advantage that the collapsing threshold can be exactly determined (see below).
Henceforth, we take $\epsilon$ and $\sigma$ as units of energy and length, respectively.
Moreover, the temperature $T$ is expressed in units of $\epsilon$ (we set Boltzmann's constant $k_B=1$).

For an isotropic potential $u(r)$ of finite strength, a sufficient condition for thermodynamic instability is $\widetilde{u}(0)<0$~\cite{Ruelle1969}, where $\widetilde{u}(k)$ is the Fourier transform of $u(r)$; on the other hand, 
if $\widetilde{u}(k)\ge 0$ for all $k$, then the system is stable~\cite{Fisher1966}.
This implies that, with
the DG potential (Eq.~\ref{eq1}), a system is thermodynamically stable for $\eta\le\eta_c=0.094031\ldots$, unstable for $\eta>\eta_c$.

We carried out Monte Carlo simulations of the DG fluid with the aim of exploring --- well beyond the usual dilute limit --- the route toward Ruelle-Fisher instability in a quantum system,
contrasting the system behavior with that of its classical counterpart.
By simulating the system beyond the collapse threshold, we also searched for indications of subextensive scaling of the emerging cluster.
Our study is completed by an analysis of the structure and superfluidity of the collapsed system, also in comparison with a liquid droplet in equilibrium with vapor.

To set the stage for subsequent analysis, it is useful to investigate first the classical DG fluid as a function of $\eta$.
We employ Gibbs-ensemble Monte Carlo (GEMC) simulations~\cite{Panagiotopoulos1987,Frenkel2002} to determine liquid-vapor coexistence and isothermal-isobaric MC simulations to locate the stability region of the triangular crystal.
To this purpose, a large-size crystal is heated isobarically until a jump is observed in the values of the number density ($\rho$) and energy ($E$).

\begin{figure}
\includegraphics[width=8.5cm]{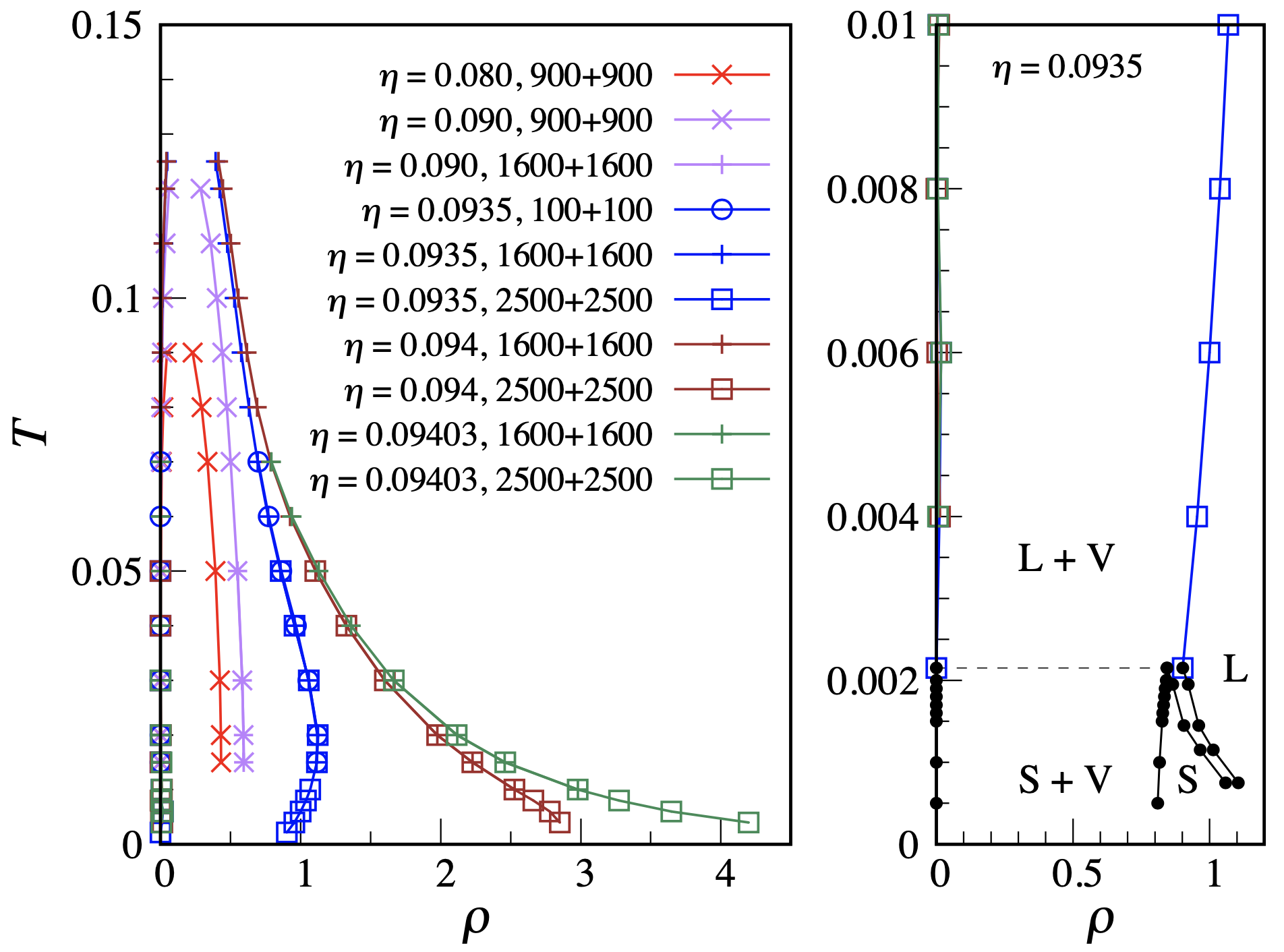}
\caption{Phase diagram of the two-dimensional classical DG model on the $\rho$-$T$ plane.
Left: Liquid-vapor coexistence data for a few values of $\eta<\eta_c$ and various choices of the initial numbers of particles in the two simulation boxes (in the legend).
Right: Magnification of the low-temperature region for $\eta=0.0935$.
The triangular solid (S) is stable in a small density window, bounded to the left by the vapor (V) and to the right by the liquid (L).
The dashed line marks the triple temperature.}
\label{fig1}
\end{figure}

Liquid-vapor coexistence points for a number of $\eta$ values close below $\eta_c$ are plotted in Fig.~1 left.
Up to $\eta\approx 0.09$ the shape of the binodal line is usual.
As $\eta$ approaches $\eta_c$, however, the coexistence region becomes wider and wider near zero temperature, thus progressively eroding the solid region (whose boundaries are less sensitive to $\eta$~\cite{Prestipino2014,Speranza2014}).
In particular, notice the substantial increase in width of the two-phase region on going from $\eta=0.094$ to $\eta=0.09403$ (a value only $3\times 10^{-5}$ higher), suggesting that the liquid density diverges as $\eta\rightarrow\eta_c$ and $T\rightarrow 0$ simultaneously.
For $\eta=0.0935$, where the density of the saturated liquid is around 1, the solid is confined to a tiny region close to $T=0$ (Fig.~1 right), which would shrink even further for larger $\eta$.
We did not observe a hexatic phase, whose existence is possible in an extremely narrow temperature range (no more than $10^{-4}$ wide) above the solid phase~\cite{Prestipino2011}, and thus tentatively assume a first-order melting transition.
We found no evidence of cluster solids, consistently with the predictions of Ref.~\onlinecite{Likos2001}.
More results for the classical DG fluid, including a survey of the structure of two-phase coexistence 
at low temperature, are presented in the Supplemental Material~\cite{supplementary} (see also references \cite{Mayer1965,Binder2012,Prestipino2015} therein).

For the simulations of the quantum system, we 
used the continuous-space worm algorithm~\cite{Boninsegni2006,Boninsegni2006b}.
If periodic boundary conditions are adopted, we compute the superfluid fraction $f_s$ using the winding-number estimator~\cite{Pollock1987};
otherwise, in a droplet regime, we employ the area estimator~\cite{Sindzingre1989,Kwon2006}. 
The relative importance of quantum effects is embodied \cite{Feynman98,Kora2020} in the parameter $\Lambda=\hbar^2/(m\epsilon\sigma^2)$, $m$ being the particle mass.
We find that, for $\eta=0.09$ and $\Lambda=0.02$, the pair correlation function (PCF) at a density $\rho\sim 1$ and at temperature $T\sim 0.1$ is nearly indistinguishable from that of the classical system.
Unless otherwise specified, we use this value of $\Lambda$ in our simulations (we also present, for comparison, some results obtained with $\Lambda=0.04$).

A sample of raw simulation data for $\eta=0.0935$ is reported in Fig.~2.
Here, we plot the pressure $P$ as a function of $\rho$ at $T=0.08$; as evidenced by the two van der Waals loops in $P(\rho)$, upon compression the 
vapor 
undergoes two first-order phase transitions.
The liquid-like character of the two denser phases, referred to as L1 and L2, is demonstrated by the PCF (see Fig.~2 inset), which is less structured for the phase of higher density (L2).

\begin{figure}
\includegraphics[width=8.5cm]{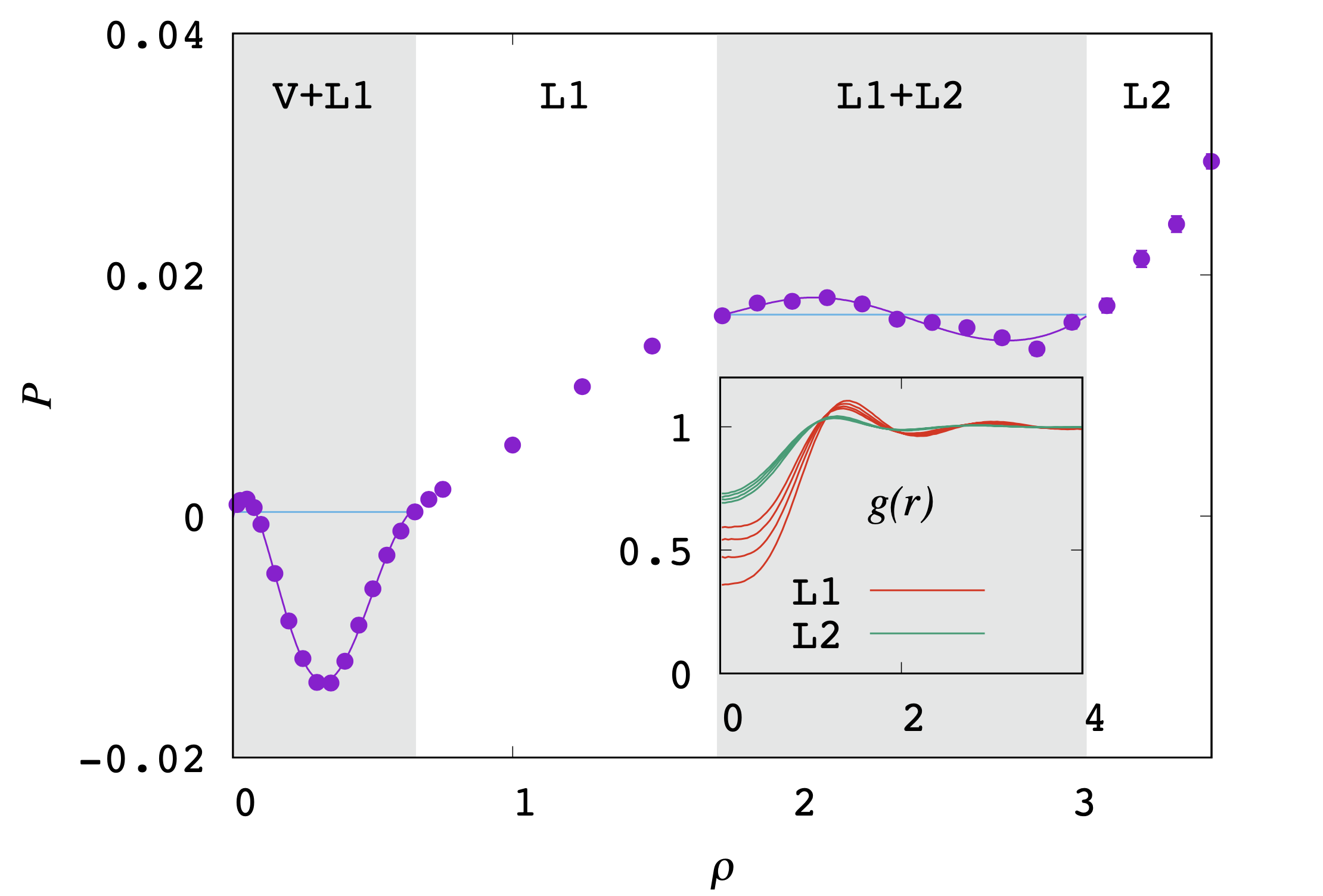}
\caption{Pressure versus density for the quantum DG model with $\eta=0.0935$ at $T=0.08$.
The grey bands are phase-coexistence regions, located by applying the equal-area rule to the $P$ vs $1/\rho$ curve.
Inset: PCF of L1 and L2 at a few densities (between 1 and 1.75 for L1; between 3 and 3.75 for L2).}
\label{fig2}
\end{figure}

\begin{figure}
\includegraphics[width=9cm]{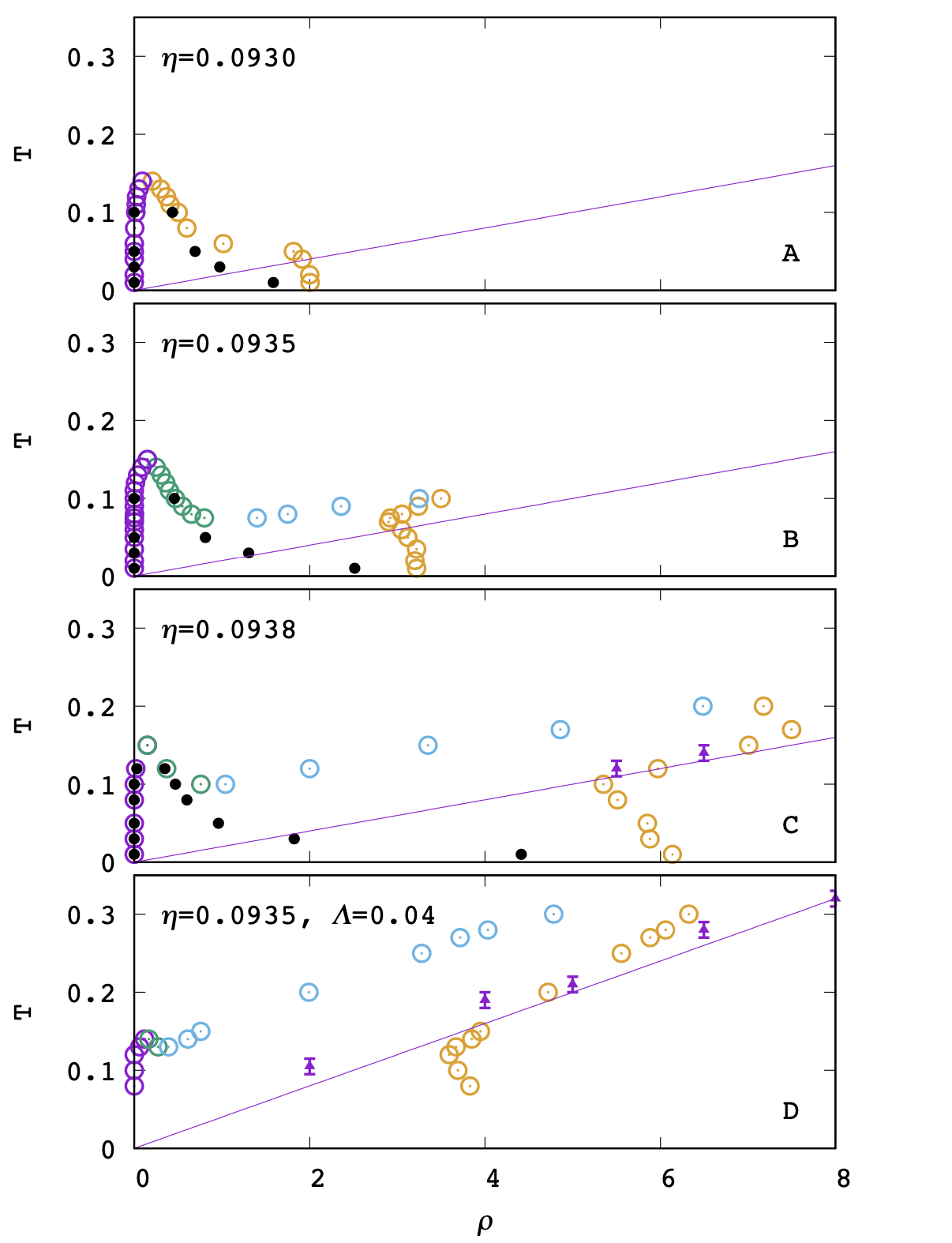}
\caption{Phase diagram of the two-dimensional quantum DG model for $\Lambda=0.02$ and three values of $\eta$:
0.093 (A), 0.0935 (B) and 0.0938 (C).
In panel D, $\Lambda=0.04$ and $\eta=0.0935$.
The circles are coexistence points which are colored differently depending on the phase and the transition involved (see text).
Black dots refer to distinguishable particles.
Also shown are the straight lines ${T}=\Lambda \rho$ (see text).
A few points on the BKT line are plotted as triangles.
The solid phase, if it exists at all, would only be stable at temperatures lower than 0.01.}
\label{fig3}
\end{figure}

Figure 3 shows the computed phase diagram of the quantum DG model for three values of $\eta$ (0.093, 0.0935 and 0.0938) with $\Lambda=0.02$, and for $\eta=0.0935$ with $\Lambda=0.04$.
The phase behavior of distinguishable quantum particles (black dots in Fig.~3) mimics that of classical particles.
On the other hand, the Bose system displays a much richer phase diagram.
Far away from $\eta_c$ (panel A) the only phase transition present is between vapor and liquid, with the liquid becoming superfluid below the Berezinskii-Kosterlitz-Thouless (BKT) line, very much like, e.g., two-dimensional $^4$He~\cite{Ceperley1989,Gordillo1998}.
Close to $\eta_c$, a second liquid phase appears (L2), which coexists with the low-density liquid (L1) between a temperature $T_t$ and an $\eta$-dependent critical temperature $T_c$ (panels B and C).
Below $T_t$, the L2 phase instead coexists with the vapor, and a cusp on the saturated L2 line at the triple temperature $T_t$ marks this change.
The L2 phase acquires superfluid properties below the BKT line, which hits the saturated L2 line close to $T_t$ (we have to say more on this later).
As $\eta_c$ is approached more and more closely, the L2-L1 and L2-vapor regions become increasingly wider, pushing the entire saturated L2 line toward higher densities.
This feature is enhanced in a system featuring more significant quantum effects, i.e., characterized by a greater value of $\Lambda$ (panel D).
An analysis of the structure and superfluidity of L2 droplets in vapor is reported in \cite{supplementary}.
Furthermore, we found no trace of cluster solids, in accordance with numerical simulations of Gaussian-core bosons in 2D~\cite{Kroiss2016}.

A first-order LLPT is unusual for one-component Bose fluids with isotropic interaction, if not even novel.
It is an exquisitely quantum phenomenon made possible by particle indistinguishability and favored (at temperatures slightly larger than $T_t$) by the disparity in structure between a normal liquid of non-overlapping particles (L1, $\rho\approx 1$) and a denser, almost structureless liquid (L2).
As long as cluster crystals are absent, we expect that a LLPT will be a generic occurrence in a Bose fluid near collapse.

Next, we explore the quantum nature of the L2 liquid as a function of temperature, by computing the superfluid fraction along a number of isochores.
Typical results are plotted in Fig.~4, reporting $f_s$ data for a few sizes in the case $\eta=0.0935,\Lambda=0.04$ and $\rho=4$.
The standard way of estimating the superfluid transition temperature $T_{\rm BKT}$ is by using the BKT recursive relations (see, for instance, Ref.~\onlinecite{Ceperley1989}).
In two different cases (corresponding to the two lower panels in Fig.~3) we find that the BKT line is well approximated by the expression $T_{BKT}\approx\Lambda\rho$ (reduced units), which is based on the well-known ``universal jump'' condition \cite{Nelson1977} and has been shown to afford quantitatively accurate predictions of $T_{BKT}$ in rather different 2D Bose systems, even in the presence of long-ranged interactions \cite{Filinov2010,Zhang2023}.
This criterion suggests the BKT line will in some cases intersect the saturated L2 line above $T_t$, e.g., for $\eta=0.0935$ and $\Lambda=0.04$, where in a range of temperatures above $T_t$ the DG model would then exhibit a first-order liquid-to-superfluid transition.
This exciting finding raises the prospect of a coexistence between normal-liquid and superfluid states, which, to our knowledge, has hitherto never been observed in a (real or model) quantum fluid.

\begin{figure}
\includegraphics[width=8.5cm,height=6cm]{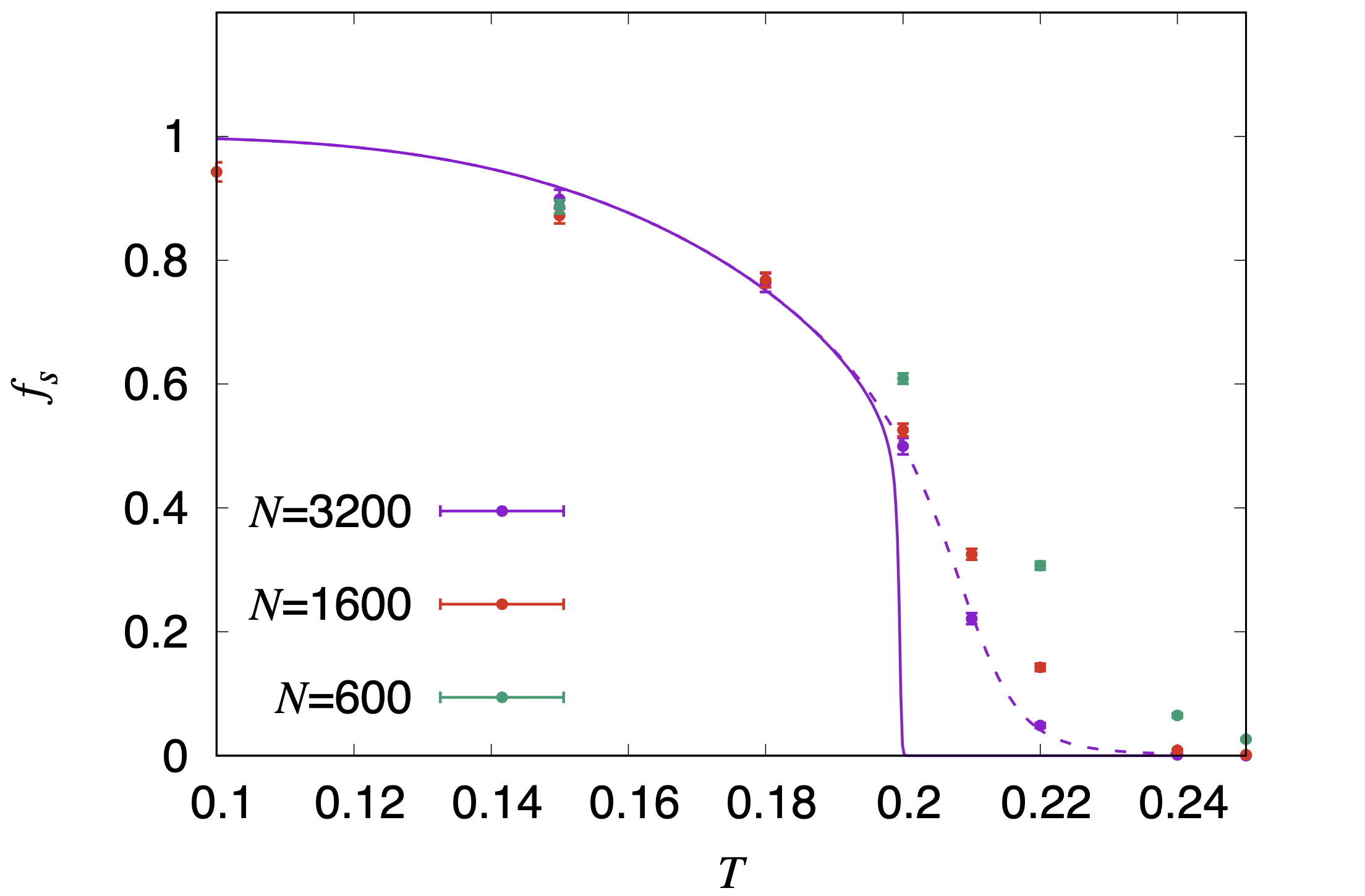}
\caption{Superfluid fraction vs temperature for $\eta=0.0935,\Lambda=0.04$ and $\rho=4$.
Data for various sizes are compared (see legend).
The dashed line is a fit through the data for the biggest size.
The continuous line is an extrapolation to infinite size based on Kosterlitz-Thouless theory (see e.g. Ref.~\cite{Ceperley1989}).}
\label{fig4}
\end{figure}

As for the behavior of the model in the unstable regime, we consider the case $\eta=0.2$.
For the classical model, at $T=1$ or larger we find the same phenomenology described in Ref.~\onlinecite{Malescio2015}, that is, the existence
of a characteristic density $\rho_\times(T)$, increasing with $T$, marking the crossover from a region of full-blown instability on the high-density side, where the collapse of a fluid sample occurs very fast, to a region of apparent stability, where the system remains homogeneous for times longer than the duration of the simulation (see more in \cite{supplementary}).
Considering then the quantum DG fluid at $T=0.1$,
collapse of the sample into a compact cluster invariably occurs, even at a density as low as 0.0001.
The final cluster looks indistinguishable from a circular droplet in equilibrium, were it not for the scaling of its size and energy with $N$: while the area of ordinary droplets is an extensive property, the radius of the cluster emerging from the decay of an unstable fluid is almost independent of $N$; simply, the cluster grows in density when $N$ is increased, while its total energy scales as $N^2$~\cite{supplementary}.

In conclusion, we consider a strongly-interacting, two-dimensional Bose fluid in the proximity of collapse.
To achieve this, the interparticle potential must be finite at the origin and have an attractive component of adequate strength.
To prevent the occurrence of cluster crystals at high density we assume a DG interaction (as representative of a broader class of potentials with the same characteristics~\cite{supplementary}).
We find that the phase behavior of the nearly unstable system is unusual, in that it undergoes an unprecedented type of LLPT.
The denser liquid (L2) becomes superfluid when cooled below the BKT line;
on the other hand, the low-density liquid (L1) is only stable above the BKT line, therefore the BKT transition only occurs for L2.
As the interaction potential is tuned toward the stability threshold, the L2 phase is shifted to higher and higher densities.
Finally, the physics of the DG fluid can be observed in a system of ultracold bosons weakly dressed with a Rydberg state; specifically, the atoms should be tailored, in a range of distances well beyond the atomic diameter, with a $Q^+$-type repulsion~\cite{Likos2001} and a short-range attraction of generic shape.
In this case, stability is recovered, but in an interval of densities below freezing the liquid phase will exhibit essentially the same features present in the DG fluid.
We look forward to seeing this scenario realized in future cold-atom platforms.

F.~C. acknowledges financial support from PNRR MUR Project No.~PE0000023-NQSTI.
M.~B. acknowledges support from the Natural Sciences and Engineering Research Council of Canada.

\bibliography{arXiv}
\end{document}